# Energy-Efficient Photonic Memory Based on Electrically Programmable Embedded III-V/Si Memristors: Switches and Filters

Stanley Cheung[1,*], Bassem Tossoun, Yuan Yuan, Yiwei Peng, Yingtao Hu, Geza Kurczveil, Di Liang, and Raymond G. Beausoleil

[1]Hewlett Packard Enterprise, Large-Scale Integrated Photonics Lab, Milpitas, CA. 95035, USA
*stanley.cheung@hpe.com
‡these authors contributed equally to this work

## Abstract

We demonstrate non-volatile optical functionality by embedding multi-layer $HfO_2/Al_2O_3$ memristors with III-V/Si photonics. The wafer-bonded III-V/Si memristor facilitates non-volatile optical functionality for a variety of devices such as Mach-Zehnder Interferometers (MZIs), and (de-)interleaver filters. The MZI optical memristor exhibits non-volatile optical phase shifts $> \pi$ ($\Delta n_g^{III-V/Si} = 2.70 \times 10^{-3}$) with ~ 30 dB extinction ratio while consuming 0 electrical power consumption in a true "set-and-forget" operation. We demonstrate 6 non-volatile states with each state capable of 4 Gbps modulation. III-V/Si (de-)interleavers were also demonstrated to exhibit memristive non-volatile passband transformation with full set/reset states. Time duration tests were performed on all devices and indicated non-volatility up to 24 hours and most likely beyond. To the best of our knowledge, we have demonstrated for the first time, non-volatile III-V/Si optical memristors with the largest electric-field driven phase shifts and reconfigurable filters with the lowest power consumption.

## Introduction

Over the past few decades, processor performance has scaled accordingly to Moore's Law, however, there remains a fundamental limit in current computer architectures: the von-Neumann bottleneck [1,2]. This inherently places a limit on the amount of data that can be transferred from memory to processor. In 2008, Hewlett Packard Labs offered a potential solution towards non-volatile in-memory computing that can surpass the limitations of current von-Neumann designs [3,4]. These devices known as memristors exhibit hysteretic current-voltage (I-V) behavior which enables multi-bit non-volatile resistance states [5]. Memristors have thus emerged as a leading candidate for implementing analog based neuromorphic computing systems in the pursuit of mimicking/harnessing the behavior of mammalian brains [3,5–8]. These two-terminal devices allow a high degree of integration density in the form of nm-sized crossbar arrays, thus yielding energy-efficient and parallelized in-memory computing where data exchange between memory and a central processing unit is uninhibited [9–11]. More recently, there have been a few optical memristor demonstrations which fall under three fundamental mechanisms [12]: 1) the phase transitions, 2) valency change, and 3) electrochemical metallization. The phase transition effect is due to the transformation of an insulating material into one with metallic properties and is driven by heat [13]. The valency effect involves oxygen vacancy formation in transition metal oxides ($HfO_2$, $Al_2O_3$, $TiO_2$, etc.) thus providing a conductive pathway between two electrodes [14–17]. The electrochemical metallization effect is based on the formation of a conductive filament composed of metal ions [18]. These three fundamental mechanisms can be further classified into two groups defined by their filamentary memristive opto-electronic functionality: 1) non-volatile phase shifters, 2) and non-volatile absorbers [18]. The work described here falls under non-volatile phase shifters where a number of electro-optical interactions happen and are not necessarily independent. For instance, the oxygen vacancy filamentation affects the optical refractive index via electrical conduction, yet charge traps can also occur [19–21]. In addition, the heat generated in nano-scale filamentary regions may morph amorphous transition metal oxides into polycrystalline or crystalline states [22]. Experimentally, it is difficult to separate these mechanisms, but experimental evidence in this paper suggests significant optical phase shifts occur through the formation of a conductive pathway via oxygen vacancies ($VO^{2+}$). Recent demonstrations of waveguide integrated optical memristive switches include ITO based latching switches [23], ZnO based reflectors [14], and Ag/a-Si/Si plasmonic absorbers [18]. Recently, we have leveraged our heterogeneous III-V/Si optical interconnect platform [36,64] to integrate memristors based on semiconductor-insulator-semiconductor capacitors (SISCAP). This platform is suitable for seamless integration of quantum dot (QD) comb lasers [24–27], III-V/Si SISCAP ring modulators [28–30], Si-Ge avalanche photodetectors (APDs) [31–34], QD APDs [35,36], in-situ III-V/Si light monitors [37,38], III-V/Si SISCAP optical filters [39,40], and non-volatile phase shifters [15,16,16,17,17,41–46], which are all essential towards realizing a fully integrated optical chip. These memristors are defined by the semiconductor-oxide interface and act as non-volatile phase shifters due to a multitude of effects described previously. The benefits of co-integrating silicon photonics and non-volatile memristors provides an attractive path towards eliminating the von-Neumann bottleneck. In addition, the memristive optical non-volatility allows post-fabrication error correction for phase sensitive silicon photonic devices while consuming zero power (supplementary note 1). As a result, we believe photonic memristors can contribute to energy efficient, non-volatile large scale integrated photonics such as: neuromorphic/brain inspired optical networks [47–56], optical switching fabrics for tele/data-communications [57,58], optical phase arrays [59,60], quantum networks, and future optical computing architectures.

## III-V/Si SISCAP Memristors

The III-V/Si SISCAP memristor (Fig. 1a-c) is comprised of 300nm thick p-type Si doped at $5 \times 10^{17}$ cm$^{-3}$, alternating layers of $HfO_2/Al_2O_3$, and 150nm thick n-type GaAs doped at $3 \times 10^{18}$ cm$^{-3}$. We chose a multi-layer $HfO_2/Al_2O_3$ stack because Mahata et al. and Park et al., have shown improved resistive switching due to atomic inter-diffusion and promotion of oxygen vacancies ($VO^{2+}$) at the $HfO_2/Al_2O_3$ (HfAlO) interface [61,62]. Fig. 1d shows the energy-dispersive X-ray spectroscopy (EDS) compositional mapping and indicates confirmation of the $HfO_2/Al_2O_3$ stack as well as the HfAlO interface. Our previous attempts with pure $Al_2O_3$ yielded unstable and chaotic switching, therefore the inclusion of multi-layer $HfO_2/Al_2O_3$ has helped. During the "set" process, $VO^{2+}$ forms in both $HfO_2$, $Al_2O_3$, and at the inter-diffused $HfO_2/Al_2O_3$ (HfAlO) interface as shown in Fig. 1b and initiates a conductive path for electrons to flow. This turns the high impedance capacitor into a device exhibiting a low resistance state. During the "reset" process only the interfacial filament is believed to rupture first due to $Al_2O_3$ having less $VO^{2+}$ than $HfO_2$ [61,62], thus breaking the conductive path via a combination of Joule heating and field effect. This effectively restores the memristor in its high resistance state. We first evaluate the multi-layer memristor device electrically with a 125 μm wide capacitive structure shown in Fig. 1e. The voltage is first swept from 0 to -10 V with a compliance current = 0.5 μA which initiates the $VO^{2+}$ electro-forming process as shown in Fig. 1f. Next, a series of voltage sweeps were performed to achieve set/reset states for multiple cycles. Given the large device surface area and random electro-formation [22,63], consecutive set cycles were observed to increase the "set" current, (Fig. 1f) indicating increased filamentation sites. We did not further explore possible bias conditions to minimize this effect given limited test structures.

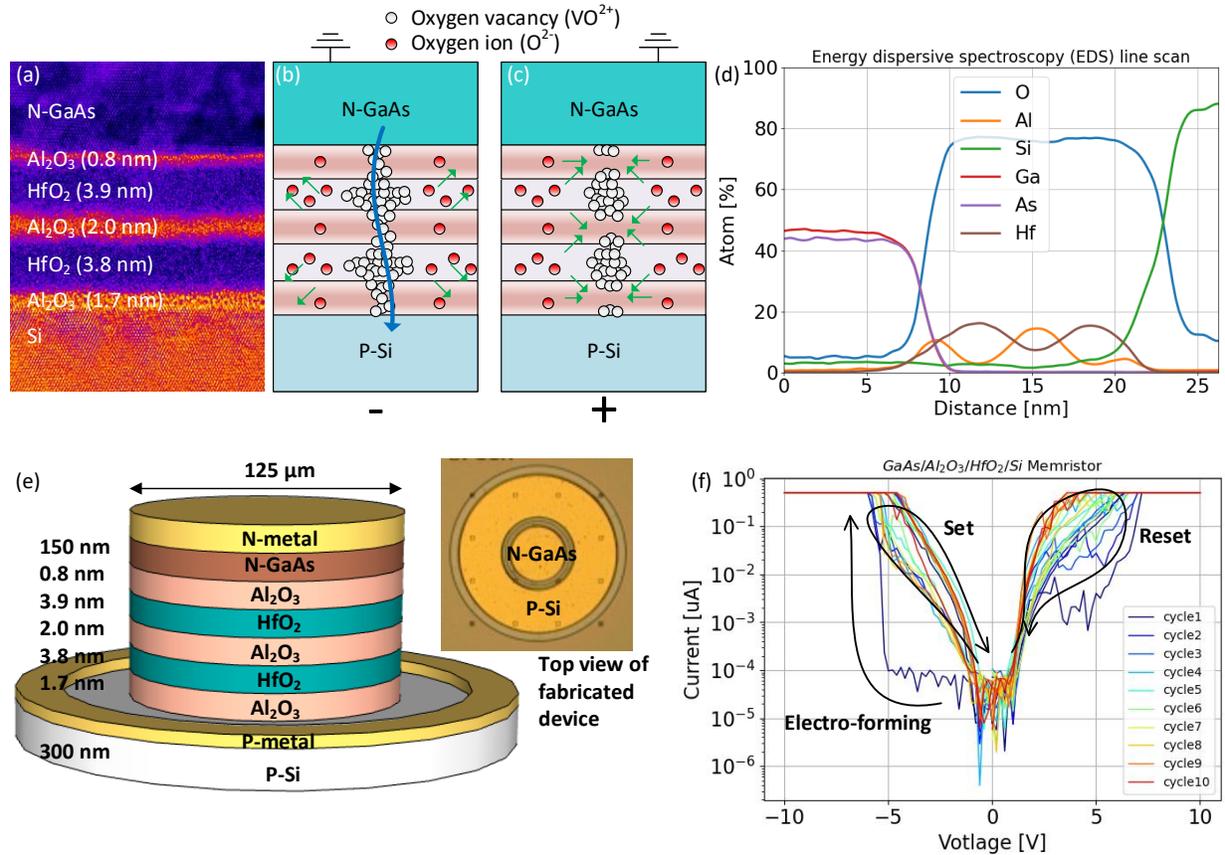

Fig. 1. (a) HRTEM image of III-V/Si SISCAP memristor stack, (b) "set" process: oxygen vacancy ($VO^{2+}$) formation which initiates conductive filamentation, (c) "reset" process: break-up of filamentation, and (d) EDS line scan for atomic composition, (e) 3-D schematic of test structure, and (f) I-V curves of electro-forming, set, and reset processes

### III-V/Si Photonic SISCAP Memristors: Mach-Zehnder Interferometers (MZI)

The single-mode waveguide structure is defined by a width, height, and etch depth of 500, 300, and 170 nm respectively as indicated in Fig. *2*a. Similar to the test capacitor structure, the SOI wafer was blanket p-type doped at $5\times10^{17}$ cm$^{-3}$. The wafer-bonded III-V region is primarily 150 nm-thick n-GaAs doped at $3\times10^{18}$ cm$^{-3}$. Fig. *2*a shows the simulated transverse electric (TE) of the optical memristor. Assuming dielectric thicknesses of 1.7/3.8/2.0/3.9/0.8 nm (Fig. *2*b) and refractive indices of 1.75/1.90/1.75/1.90/1.75 for $Al_2O_3/HfO_2/Al_2O_3/HfO_2/Al_2O_3$ respectively, the calculated optical confinement factors are $\Gamma_{Si}$ = 64.49 %, $\Gamma_{HfO2}$ = 1.637 %, and $\Gamma_{Al2O3}$ = 0.82 % with an overall effective index of $n_{eff}$ = 3.0971 and group index of $n_g$ = 3.7914. For comparison, a pure silicon waveguide with oxide cladding has an effective index of $n_{eff}$ = 2.9774 and group index of $n_g$ = 3.9765. An $Al_2O_3$ layer is inserted in between the $HfO_2$, because it was experimentally determined to be easier to wafer-bond $Al_2O_3$ to $Al_2O_3$ rather than $HfO_2$. The choice of n-GaAs over p-GaAs was two-fold: 1) lower optical absorption loss from dopants, and 2) easier III-V/Si laser integration. Also, GaAs exhibits ~ 4 × smaller electron effective mass and ~ 6 × larger electron mobility ($m_e^*$ = 0.063$m_0$, $\mu_e$ = 8500 cm$^2$/V-s) than crystalline Si ($m_e^*$ = 0.28$m_0$, $\mu_e$ = 1400 cm$^2$/V-s) [36,40,64]. Therefore, the plasma dispersion effect on index change in n-type GaAs is more efficient with lower free carrier absorption (FCA) loss.

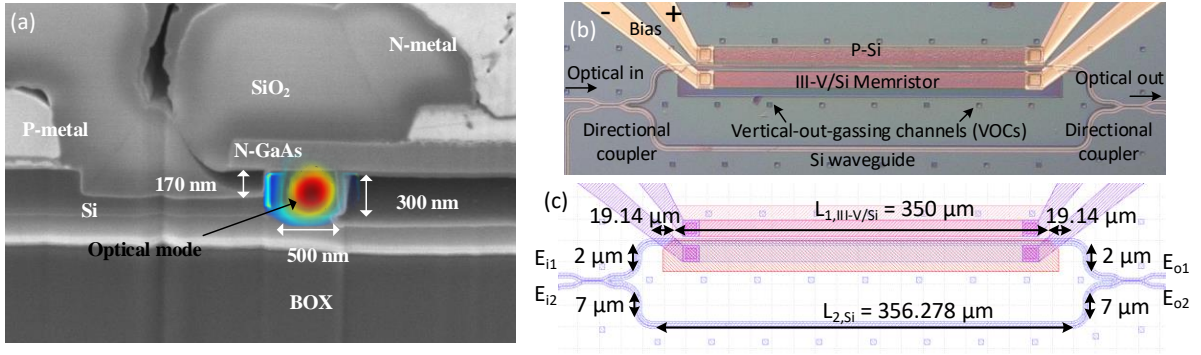

Fig. 2. (a) SEM cross section of optical memristor with simulated guided optical mode. (b) HRTEM image of memristor dielectric stack, and (c) EDS line scan for atomic composition.

The electric field transmission function of the top/bottom arm of the MZI ($E_{o1}$, $E_{o2}$) in Fig. 2c can be modeled with the following transfer matrix:

$$\begin{bmatrix} E_{o1} \\ E_{o2} \end{bmatrix} = \begin{bmatrix} t & jr \\ jr & t \end{bmatrix} \begin{bmatrix} e^{i\phi_1 - \zeta_1} & 0 \\ 0 & e^{i\phi_2 - \zeta_2} \end{bmatrix} \begin{bmatrix} t & jr \\ jr & t \end{bmatrix} \begin{bmatrix} E_{i1} \\ X_{i2} \end{bmatrix} \quad (1)$$

$$I_{o1} = |E_{o1}|^2, \quad I_{o2} = |E_{o2}|^2 \quad (2)$$

$$\phi_1 = \beta_{1,si} L_{1,si} + \beta_{1,III-V/Si} L_{1,III-V/Si}, \quad \phi_2 = \beta_{2,si} L_{2,si}$$
$$\zeta_1 = \frac{1}{2}\alpha_{1,si} L_{1,si} - \frac{1}{2}\alpha_{1,iii-v/si} L_{1,III-V/Si}, \quad \zeta_2 = \frac{1}{2}\alpha_{2,si} L_{2,si} \quad (3)$$

The variables $\beta_{1,si}$, $\beta_{2,si}$, $\beta_{1,iii-v/si}$, represent the propagation constants of the top arm silicon, bottom arm silicon, and top arm III-V/Si memristor waveguide respectively. $\alpha_{1,si}$, $\alpha_{2,si}$, $\alpha_{1,iii-v/si}$ represent the optical losses in the top arm, bottom arm, and top arm III-V memristor waveguide respectively. $L_{1,si}$, $L_{2,si}$, $L_{1,iii-v/si}$ represent the corresponding lengths. Based on the transfer-matrix model, the two directional couplers have a power transfer coefficient of 49 % assuming they are identical during fabrication. The measured spectrum indicates a free spectral range (FSR) of ~ 20.13 nm with an extinction ratio (ER) of ~ 31.1 dB near 1310 nm. The III-V/Si memristor region is located on the upper arm with a length of $L_{1,III-V/Si}$ = 350 µm. The p-doping is defined 2.0 µm away from the edge of the silicon waveguide and test structures indicated little to no effect on optical losses. The n-GaAs has a 200 nm overhang to the edge of the silicon waveguide such that III-V/Si bonding remains intact while avoiding contact with silicon pillars as shown in Fig. 2a. A fully etched deep trench is defined in between the MZI arms such that the p-Si is electrically isolated from the wafer-bonded n-GaAs region. The III-V/Si SISCAP structure operates as the memristor. In order to investigate non-volatile optical memory functionality, we first measure the current-voltage (I-V) relationship as shown in Fig. 3a. By voltage cycling from 0 → - 21 → 0 → 15 → 0 V, a hysteresis curve is observed, therefore, confirming electrical memristor behavior. Fig. 3b. illustrates the corresponding resistance indicating an initial high-resistance-state (HRS) which becomes a low-resistance-state (LRS) by applying a set voltage $V_{set}$ = - 17.31 V.

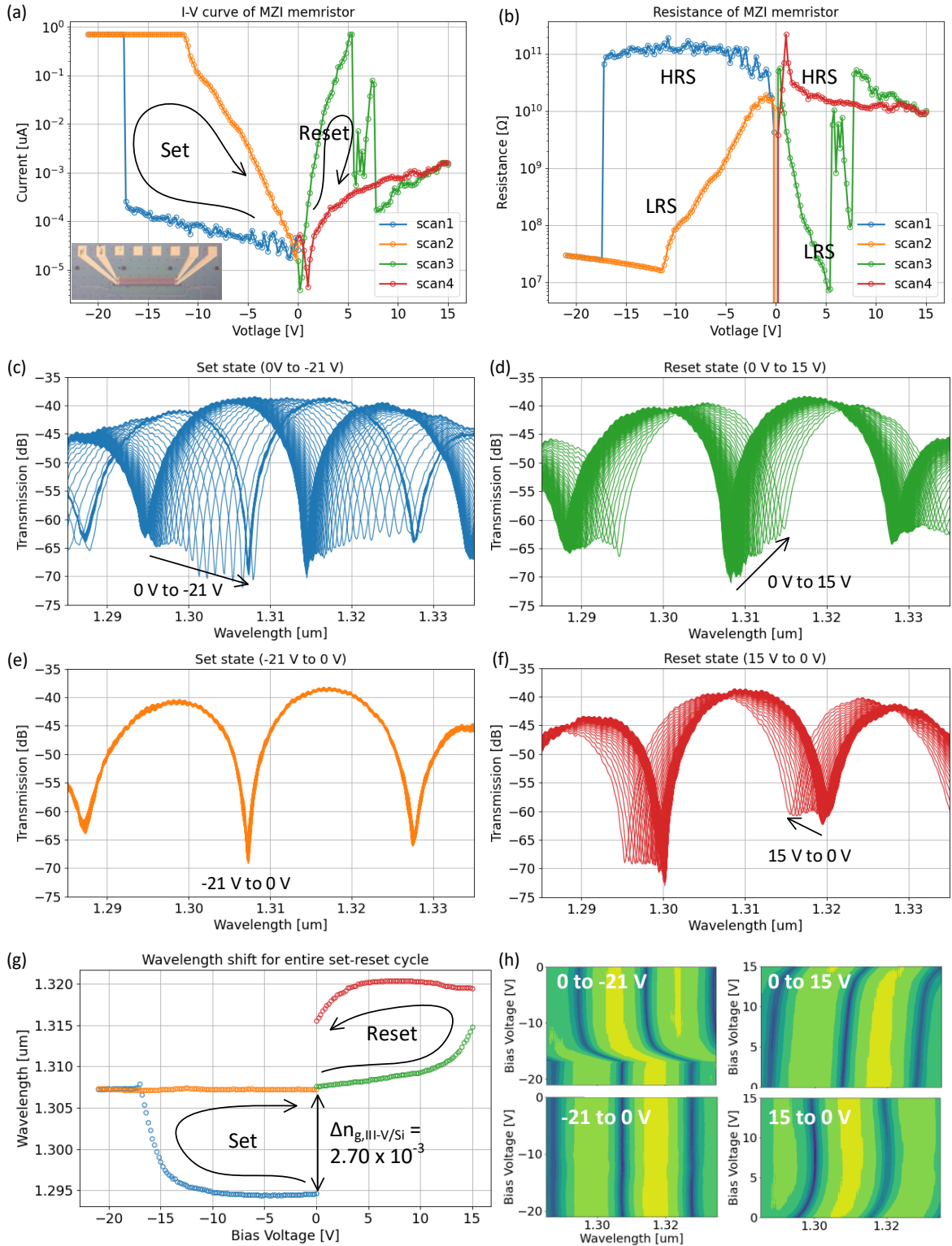

Fig. 3. (a) Measured I-V hysteresis indicating non-volatile memristive set/reset states. (b) Corresponding resistance indicating regions of high-resistance-states (HRS) and low-resistance-states (LRS). Measured optical spectrum for (c) "set" (0 to -21 V), (d) turning off "set" (-21 to 0V), (e) "reset" (0 to 15 V), and (f) turning off "reset" (15 to 0 V). (g) Tracked resonance vs. voltage, (h) spectral evolution vs. voltage.

By applying a reset voltage of $V_{reset} > 5$ V, a transition from the LRS to HRS can occur, thus concluding a reset back to the original electrical state. While taking I-V data, we simultaneously measured the optical spectral response with an optical spectrum analyzer (OSA). The optical response is shown in Fig. *3*c-f and is color-coded according to the I-V curves in Fig. *3*a-b. Applying a bias from 0 to -21 V results in a non-volatile wavelength shift of $\Delta\lambda_{non-volatile} = 12.53$ nm wavelength shift with near negligible optical losses (Fig. *3*c). Ramping back down from -21 to 0 V does not shift the optical response back to the original state (Fig. *3*d) and has a non-volatile wavelength stability of ~ +/- 0.2 nm (35 GHz). This indicates a non-volatile phase shift of $\Delta\varphi = 1.245\pi$ assuming a FSR = 20.13 nm, at essentially 0 power consumption (recorded current = 34 pA at 0 V). The group index ($n_g^{III-V/Si}$) of the III-V/Si memristor can be calculated by the following: $FSR_{MZI} = \lambda^2/(n_g^{lower} L^{lower} - n_g^{upper} L^{upper})$ where $n_g^{lower} L^{lower} - n_g^{upper} L^{upper} = n_g^{Si} L_{Si}^{lower} - (n_g^{III-V/Si} L^{III-V/Si} + n_g^{Si} L_{Si}^{upper})$. From this, the group index difference was calculated to be $\Delta n_g^{III-V/Si} = 2.70 \times 10^{-3}$ which is quite significant. On separate devices, a full $\Delta\varphi > 4\pi$ can be achieved with $\Delta n_g^{III-V/Si} = 13.7 \times 10^{-3}$ (supplementary note 5). As we attempt to reset the device from 0 → +15 → 0V, a high resistance state is achieved. This shows we can electrically reset the device, albeit with the absence of an optical reset. The disassociated coupling of electrical/optical reset may indicate the existence of residual defects from $VO^{2+}$ formation (Fig. 1c) that can attributed to long-lived charge traps [65–67]. This is most likely the case since a $\Delta n_g^{III-V/Si} = 2.70 \times 10^{-3}$ would require an untenable 20% change in the $HfO_2/Al_2O_3$ multi-layer stack, assuming there are no thickness changes. To verify the absence of any significant material degradation, we performed HRTEM imaging for the virgin, set, and reset states as shown in supplementary section 3. Geometric phase analysis (GPA) was also performed to fully quantify any strain deformation that may occur for virgin, set, and reset states. The results, shown in the supplementary section, indicates the set process contributes to an increase of nano-scaled oxide/semiconductor interfacial strains ranging from -0.5 to 0.5 % for the in-plane ($E_{xx}$) and out-of-plane ($E_{yy}$) directions. To the degree these nano-scaled strain points contribute to electron charge traps or $VO^{2+}$ is quantitatively unknown in our device, but are known to exist in other studies [67–70]. In order to quantitatively assess the charge trap density needed to observe experimental phase shifts ($\Delta n_g^{III-V/Si} = 2.70 \times 10^{-3}$), we employed SILVACO ATLAS. This is a two-dimensional solver capable of performing energy-band diagram and charge concentration calculations to theoretically predict optical effective and group index changes as a function of trapped charge density ($Q_{TC}$). Based on the electron and hole concentrations, a spatial change in index can be calculated as [71,72]: $\Delta n(x,y)$ (at 1310 nm) $= -6.2 \times 10^{-22} \Delta N(x,y) - 6 \times 10^{-18} \Delta P(x,y)^{0.8}$, where x and y are the 2D lateral and vertical dimensions as detailed in supplementary note 1. The resulting spatial indices are then used in an optical finite-difference-eigenmode (FDE) solver to calculate non-volatile group index changes $\Delta n_{g,non-volatile}$ vs. $Q_{TC}$ as indicated in supplementary note 1. A charge trap density of $Q_{TC} = 9 \times 10^{19}$ cm$^{-3}$, yields a group index change of ~ $1.75 \times 10^{-3}$ which is not too far off from our experimentally determined value of $\Delta n_g^{III-V/Si} = 2.70 \times 10^{-3}$. An extreme case of a phase shift $\Delta\varphi > 4\pi$ is demonstrated in supplementary note 5 and exhibits an index change of $\Delta n_g^{III-V/Si} = 13.7 \times 10^{-3}$ with essentially 0 static power consumption. This large change indicates residual charge trap densities $> 9 \times 10^{19}$ cm$^{-3}$. Set switching speeds were demonstrated to be ~ 1ns [15].

In order to test the reliability of the non-volatile states, time duration tests were performed by biasing the MZI memristor into multiple non-volatile states and the optical response was recorded for 24 hours every 5 minutes. In Fig. 4a, the red curve is the initial state at 0 V. Next, we bias the device to "set state 1", turn off the bias and record the optical output for 24 hours. Next, we perform the same procedure for "set state 2". As observed in Fig. 4a, the optical response in the non-volatile set states are stable up to 24 hours and most likely beyond. In order to quantify this stability, we extracted the resonance dips over time (indicated by °). As a result, the multiple set states are stable by ~ +/- 0.05 nm (8.77 GHz) within a 24 hour time frame indicating stable non-volatile behavior. The extracted non-volatile power difference between the initial state and different set states are indicated by '×' in Fig. 4b and show the possibility of multi-bit non-volatile weighting.

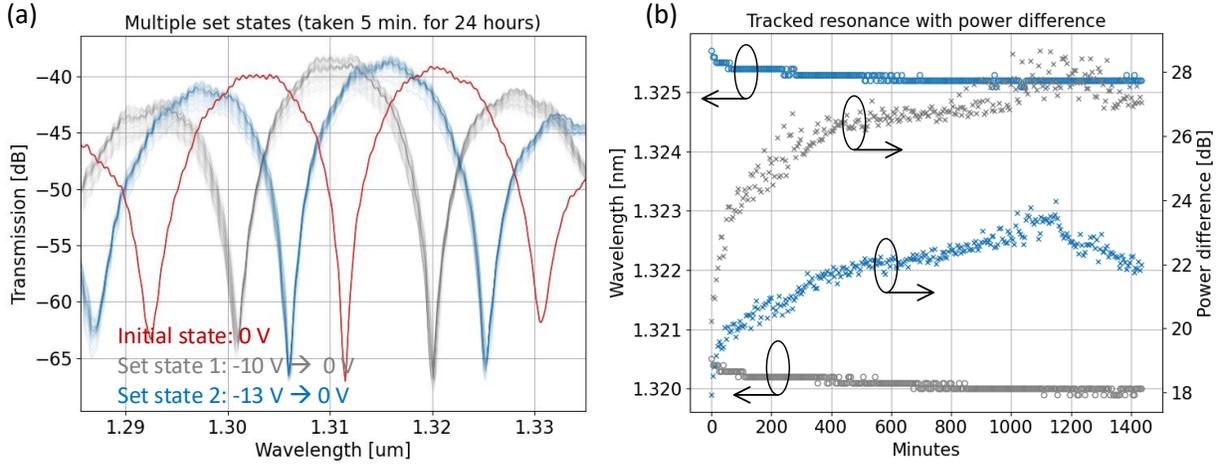

Fig. 4. (a) 24 hour measured optical response taken every 5 minutes for (a) set state 1 (gray), set state 2 (blue), (b) tracking resonance minimum and power difference of set state 1 and 2 over 24 hours.

### III-V/Si Photonic SISCAP Memristors: (De-) Interleavers

Ring assisted AMZIs (RAMZIs) find use as filters for flat-top response with improved channel XT [30,36,39,40]. They also find use as linearized transfer functions for improved bit resolution in optical neural networks [73] as well as RF photonics [74]. For the (de-)interleaver architecture, we chose a single ring resonator assisted asymmetric Mach-Zehnder interferometer (1-RAMZI) where the transmission passbands can be expressed as:

$$\Phi_{1-ring\ RAMZI} = \begin{bmatrix} c_1(\lambda) & -js_1(\lambda) \\ -js_1(\lambda) & c_1(\lambda) \end{bmatrix} \begin{bmatrix} A^R(z)/A(z) & 0 \\ 0 & e^{j2\pi n_g(\lambda)L_{ring}/\lambda} \end{bmatrix} \begin{bmatrix} c_0(\lambda) & -js_0(\lambda) \\ -js_0(\lambda) & c_0(\lambda) \end{bmatrix} \quad (4)$$

$$A^R(z) = \sqrt{1-\kappa_r} + (e^{j2\pi n_g(\lambda)L_{ring}/\lambda})^{-2}, \quad (5)$$

$$A(z) = 1 + \sqrt{1-\kappa_r}(e^{j2\pi n_g(\lambda)L_{ring}/\lambda})^{-2} \quad (6)$$

The AMZI bar and cross port transmission are respectively defined similarly for the MZI filter with the addition that the $\kappa_r$ is the ring coupling coefficient. The FSR is defined by the ring circumference such that the FSR = $c/n_g/L$. Therefore, a channel spacing of 65 GHz for the 1-ring AMZI requires $L_{ring}$ = 1200 μm for a calculated group index of $n_g$ = 3.78. The ideal ring resonator coupling for a 1-RAMZI occurs at $\kappa_r$ = 0.89. Details of this device under volatile SISCAP phase shift operation can be found in [30,36,39,40].

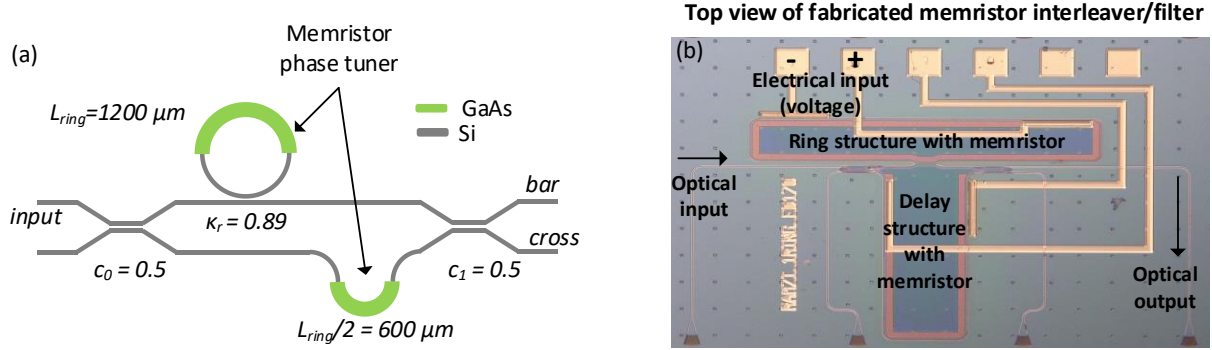

Fig. 5. (a) Schematic of 65 GHz 1-RAMZI (de-)interleaver with memristive phase tuning elements (green), and (b) top view of fabricated device.

The III-V/Si SISCAP structure on the ring or delay path can operate as the optical memristor. In order to investigate non-volatile optical memory functionality, we once again measure the current-voltage (I-V) relationship as shown in Fig. 6 (a). A hysteresis curve is observed, therefore, confirming electrical memristor behavior. While taking I-V data, we simultaneously measured the spectral response with an optical spectrum analyzer (OSA). The optical response is shown in Fig. 6 (c) – (f) and is color-coded according to the I-V curves in Fig. 6 (a). Applying a bias from 0 to -10 V and back down to 0V results in a non-volatile change in the passbands. In an attempt to reset the optical response, we apply a bias from 0 to 5 V and back down to 0V. A passband shape similar to the initial one is obtained with minor differences possibly associated with remaining $VO^{2+}$ charge traps. Transfer matrix modeling from equations (4) – (6) indicate a ring resonator phase difference of ~ $0.89\pi$ ($\Delta n_g^{III-V/Si} = 0.48 \times 10^{-3}$) from the initial to set state. It is observed that the "set" voltage (- 6 V) is much less than that of the MZI (- 17 V) and may be due to the differences in memristor area by a factor of 3.4. If these optical memristive filters were to be used as non-volatile elements, reliability and retention times would be of interest.

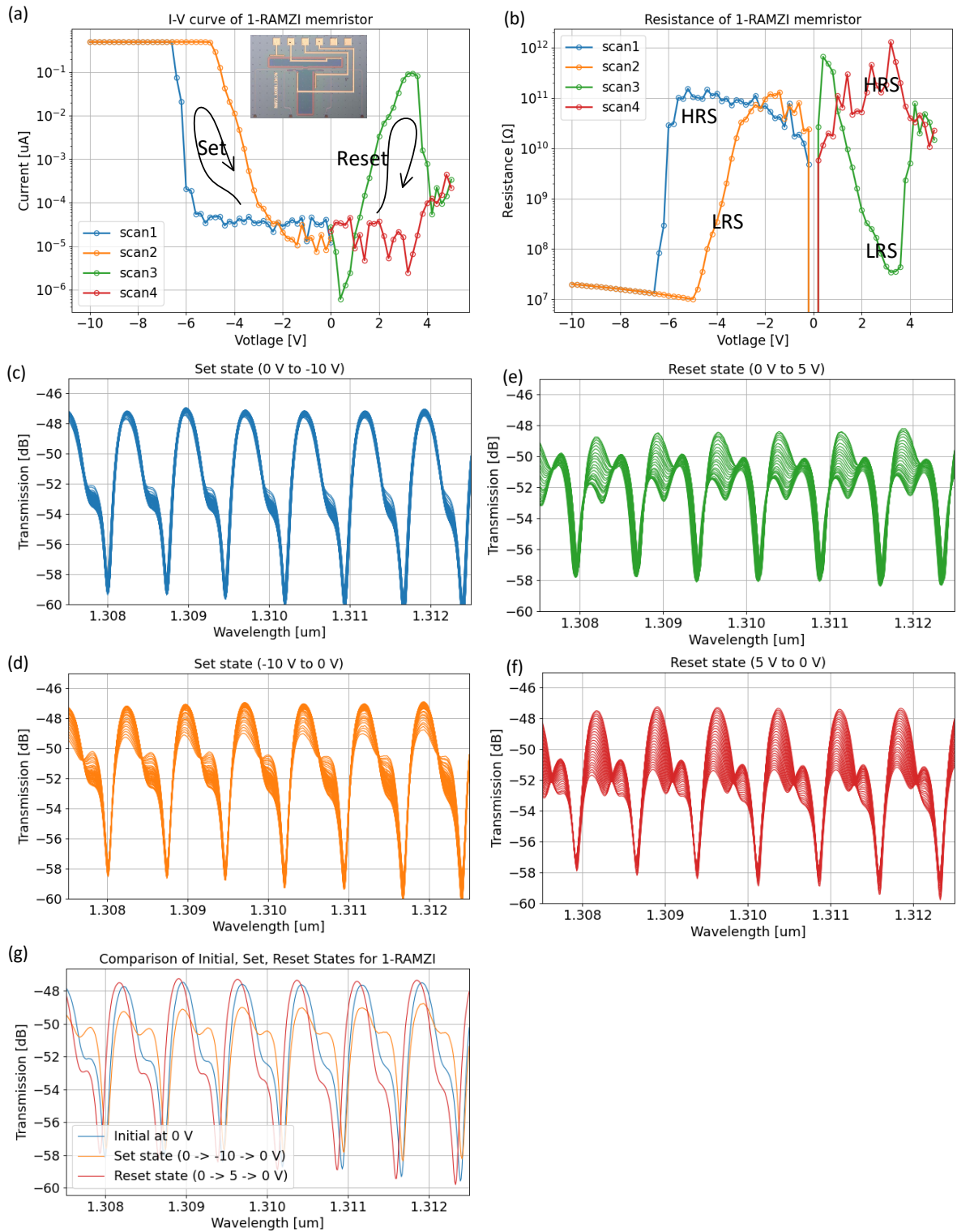

Fig. 6. (a) Measured I-V hysteresis indicating non-volatile memristive set/reset states and (b) corresponding resistance states. Measured optical spectrum for (c) "set state" (0 to -10 V), (d) turning off "set state" (-10 to 0 V), (e) "reset state" (0 to 5 V), and (f) turning off "reset state" (5 to 0V). (g) Overlapped spectrum for final initial, set, and reset states.

We performed these time duration tests on a separate (de-)interleaver known as a 3rd order AMZI. Multiple non-volatile set states were achieved with each one measured every 5 minutes for a 24 hour period. The overlapped filter shapes are shown in *Fig. 7*a. Three minima from each state were also tracked and exhibited +/- 0.02 nm (3.51 GHz) change for the worst case (blue curve) for this 24 hour period.

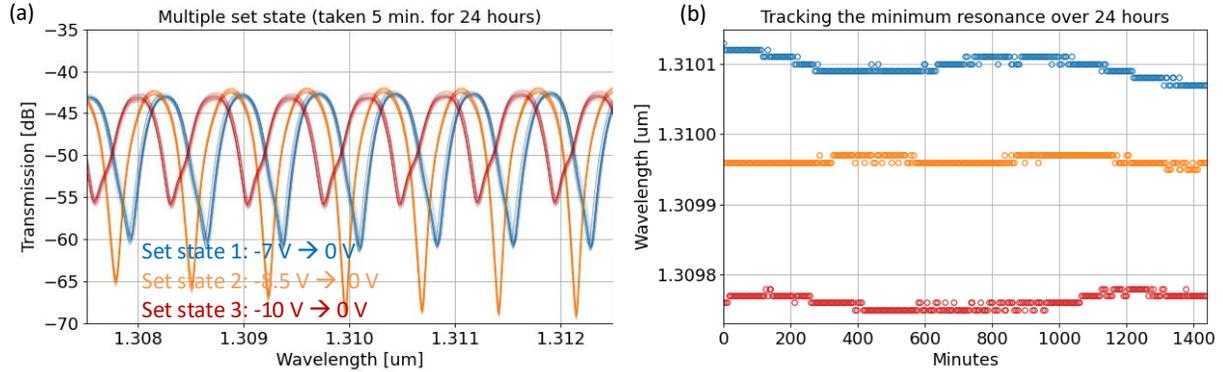

Fig. 7. (a) Multiple set states for 3rd order AMZI (de-)interleaver, and (b) non-volatile stability over a 24 hour period by tracking spectra minima.

## Heterogeneous III-V/Si photonic in-memory computing platform

Fig. *8* illustrates our vision of co-locating memristors and silicon photonics for in-memory optical computing on a common heterogeneously integrated substrate [36,64]. Fig. *8*a shows a general artificial neural network (ANN) architecture composed of an input layer, N hidden layers, and an output layer. This ANN can be realized by our III-V/Si ONN where N layers can be achieved by time re-cycling the chip. The architecture is comprised of previously demonstrated HPE devices (III-V/Si QD lasers[75], SISCAP MZI[40,76], lossless light monitors[37,38], QD APDs[35,77], Ge APDs[31–33,78], programmable nonlinear activation functions[73,79], etc.) as well as the non-volatile III-V/Si memristor (this work) that can be placed in the MZIs. The weights of an entire network can be trained by using the III-V/Si MZIs in a low-voltage, volatile push-pull operation (up to ~ 30 Gbps in traveling wave electrode configuration [80]) while keeping track of the weight amplitudes with the III-V/Si lossless light monitor [37,38]. Fig. *8*b shows a total tuning wavelength of 7.64 nm can be achieved from 0 V to -5 V and with open 4 Gbps eye diagrams. For this particular MZI length, a phase shift of $0.85\pi$ can be achieved with a power consumption of 0.63 nW (~10,000,000 × improvement over thermal heaters [47]). In push-pull configuration, the wavelength tuning is higher with equivalent drive voltage. For light monitoring, we have demonstrated internal trap mediated photo-carrier detection which induces no optical loss [37,38]. These detectors can be part of a feedback circuitry with the non-volatile III-V/Si MZI CTM cell for true in-memory optical computing. Once network training is complete, one can appropriately adjust one arm of the III-V/Si MZI using the higher voltage, optical memristive effect for non-volatile inference. Current optical non-volatile neuromorphic systems use PCM materials that require: 1) high power (4 mW) optical pulses (200 ns)[81] or 2) graphene thermal heaters in pulse operation (SET: 3 V, 100 μs, RESET: 5 V, 400 ns) to change from crystalline to amorphous and vice versa. Furthermore, these set/reset operations require a temporal separation in the seconds time range to ensure thermal relaxation[82]. Reconfiguring PCMs with optical pulses places additional scalability issues in order to arbitrarily broadcast a network of control pulses into the system. In addition, ultrafast pulsed lasers can be quite energy efficient and reduce MAC/J figure of merit. For the electrical based heating approach, the μs SET time and long enough thermal relaxation duration (2 s) between pulses can affect training throughput. Our ability to train at 10s of Gbps in low-power, volatile operation combined with reliable multi-bit non-volatile inferencing allows for increased throughput and energy efficiency which are lacking in PCM based approaches and modern day electrical MVM architectures.

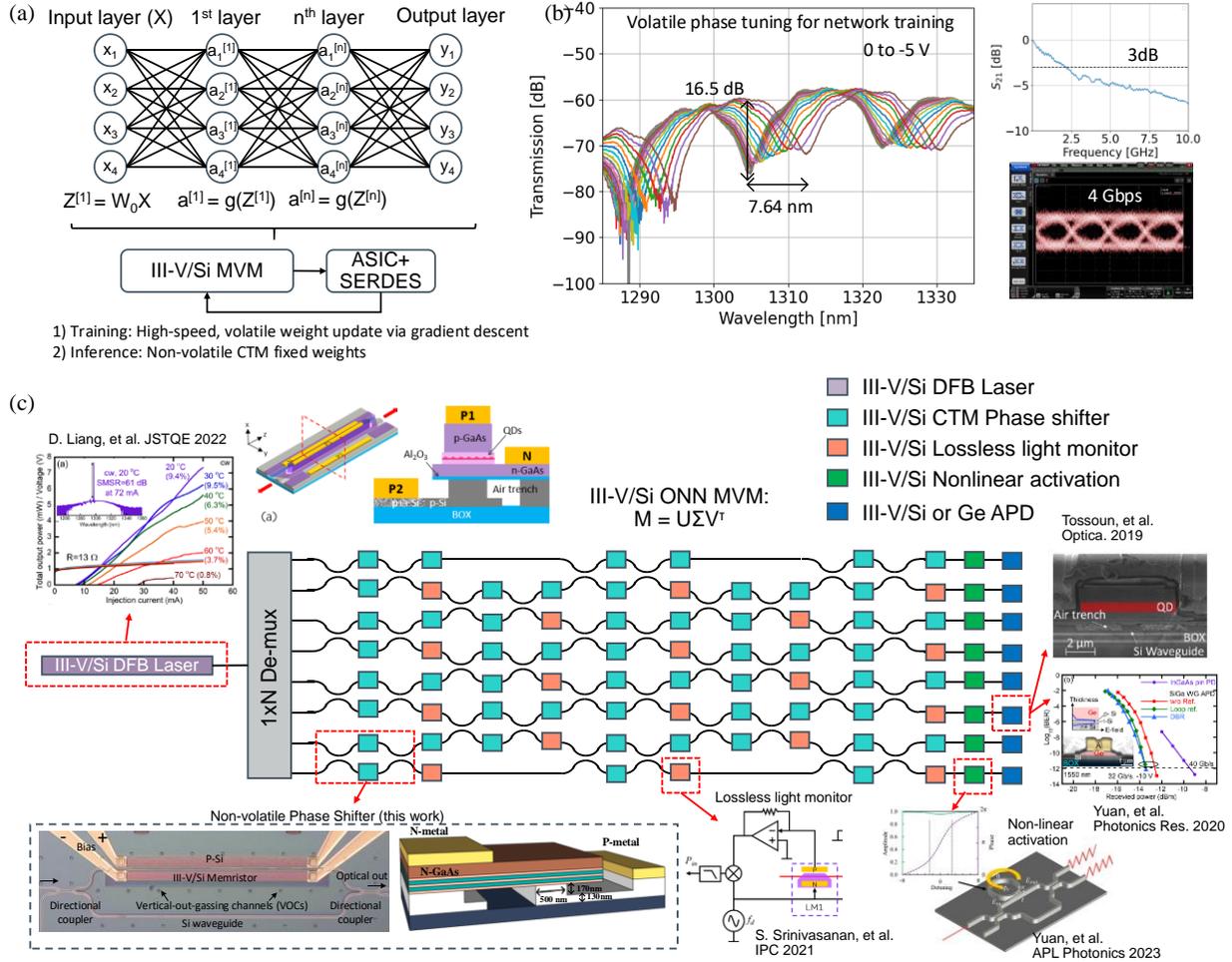

Fig. 8. Schematic of a fully-integrated ONN MVM mesh on a heterogeneous III-V/Si platform.

## Conclusion

Over the past few decades, processor performance has scaled accordingly to Moore's Law, however, there remains a fundamental limit in current computer architectures: the von-Neumann bottleneck. This inherently results in the need to transfer massive data between processor and memory with an intrinsic limit on bandwidth × distance plus increasing interconnect power consumption. As a major step towards breaking this bottleneck (especially for photonic neuromorphic computing), the work described here enables volatile operation for low-power, high-speed, on-chip training and non-volatile memristive optical memory for inference. This is all done on a heterogeneous III-V/Si platform capable of integrating all the necessary components needed for next generation applications such as: neuromorphic/brain inspired optical networks [47–55,83], optical switching fabrics for tele/data-communications [57,58], optical phase arrays [59,60], quantum networks, and future optical computing architectures. In particular, this work demonstrates for the first time, co-integration of III-V/Si memristors with optical MZI and (de-)interleaver filters which are key components in both communication and computing applications. The III-V/Si MZI memristor exhibits non-volatile optical phase shifts $> \pi$ ($\Delta n_g^{\text{III-V/Si}} = 2.70 \times 10^{-3}$) with ~ 30 dB extinction ratio while under 0 electrical power consumption. We demonstrate 6 non-volatile states with each state capable of 4 Gbps modulation. The III-V/Si (de-)interleaver memristor were also demonstrated to exhibit memristive non-volatile passband transformation with full set/reset states for 1-RAMZI and $2^{nd}$ order AMZI architectures. Time duration tests were performed on all devices and indicated non-volatility up to 24 hours and most likely beyond. In addition, the memristive optical non-volatility allows for post-fabrication error correction of phase sensitive silicon photonic devices while consuming zero power as shown in the supplementary section.

## Data availability

The data that support the findings of this study are available from the corresponding author on reasonable request.

## Competing interests

The authors declare no competing interest.

## Acknowledgements

We thank funding from DOE ARPA-E ULTRALIT contract No. DE-AR0001039, and USG MPO contract No. H98230-18-3-0001.

## Author contributions

S.C. conceived the initial concept, and designed devices. D.L. designed SISCAP structure and fabrication flow, and participated in data analysis. G.K. and Y.H. fabricated the devices and suggested improvements in the design phase. S.C., B.T., Y.Y., and Y.P. conducted the chip testing. D.L. and R.B. managed the project and gave important technical advice. All authors reviewed the manuscript.

# Supplementary

Stanley Cheung, Bassem Tossoun, Yuan Yuan, Yiwei Peng, Yingtao Hu, Geza Kurczveil, Di Liang, and Raymond G. Beausoleil

Hewlett Packard Enterprise, Large-Scale Integrated Photonics Lab, Milpitas, CA. 95035, USA
*stanley.cheung@hpe.com

## Supplementary Note 1:

### Electro-optical simulations and design

A 2-D electrical solver (SILVACO ATLAS [7]) was used to study the non-volatile electrical behavior of the III-V/Si memristor structure. This modeling software has been used successfully in other theoretical optical non-volatile structures[8,9]. The program numerically solves the Poisson, charge continuity equations, drift-diffusion transport, and quantum tunneling mechanisms. The models involved include Fermi-Dirac statistics, Shockley-Read-Hall recombination, quantum tunneling that includes direct and Fowler-Nordheim [7]. The semiconductor material parameters used for optical and electronic TCAD simulations are listed in Table *1*.

Table 1: Material parameters used in electro-optical simulations

| Material | $E_g$ (eV) | $\chi$ (eV) | VBO (eV) | CBO (eV) | n | $N_{TC}$ (cm$^{-3}$) | $\varphi_d$ (eV) |
|---|---|---|---|---|---|---|---|
| $Al_2O_3$ | 7.0 [10] | 1.5 [11–13] | 3.2 [10] | 2.7 [10] | 1.75 | - | - |
| $HfO_2$ | 5.7 [10] | 2.5 [10] | 2.7 [10] | 1.9 [10] | 1.9 | $10^{19}$ - $10^{20}$ [9,14,15] | 2.0 [16–19] |
| $SiO_2$ | 8.9 [10] | 1.3 [10] | 4.5 [10] | 3.3 [10] | 1.44 | - | - |
| Si | 1.1 [10] | 4.1 [10] | - | - | 3.507 [20] | - | - |
| GaAs | 1.43 [7] | 4.07 [7] | - | - | 3.406 [20] | - | - |

$E_g$: Energy gap, $\chi$: electron affinity, n: refractive index (at λ = 1310 nm), VBO: valance band offset, CBO: conduction band offset, $N_{TC}$: trap density, $\varphi_d$: electron trap level

Once the structure is built with the appropriately assigned models and trapped charge density, the band diagrams, C-V curves, and electron/hole concentrations are simulated for various biases. The spatial profiles of the change in electron/hole concentrations ($\Delta P(x,y)$ and $\Delta N(x,y)$) are then exported and used to calculate a spatial refractive index change $\Delta n(x,y)$ by the following equation [21,22]:

$$\Delta n(x, y) = -6.2 \times 10^{-22} \Delta N(x, y) - 6.0 \times 10^{-18} \Delta P(x, y) \qquad (1)$$

This equation is valid for a wavelength λ = 1310 nm and is used throughout the manuscript. Next, this spatial refractive index change $\Delta n(x,y)$ is exported into a 2D-FDE optical mode solver (Lumerical) and the non-volatile change in effective index ($\Delta_{neff,non-volatile}$) is calculated. This change can then be used to calculate non-volatile phase shifts on photonic devices such as ring resonators, Mach-Zehnder interferometers, filters, etc. The figure below illustrates the simulation procedure:

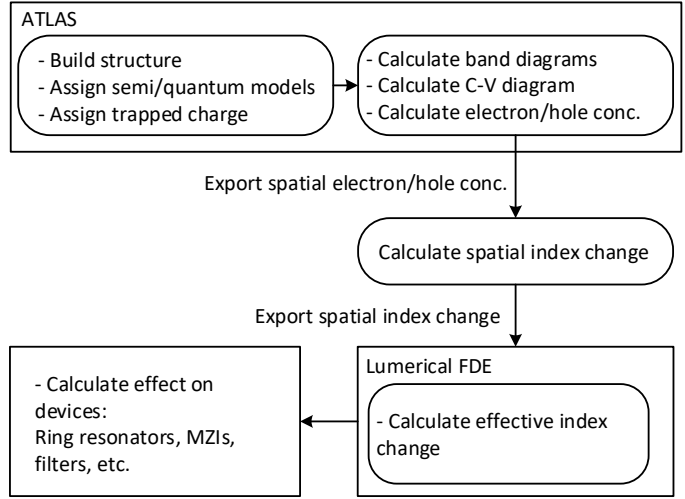

Fig. 9. Flowchart of electro-optical simulation procedure for III-V/Si memristor structure

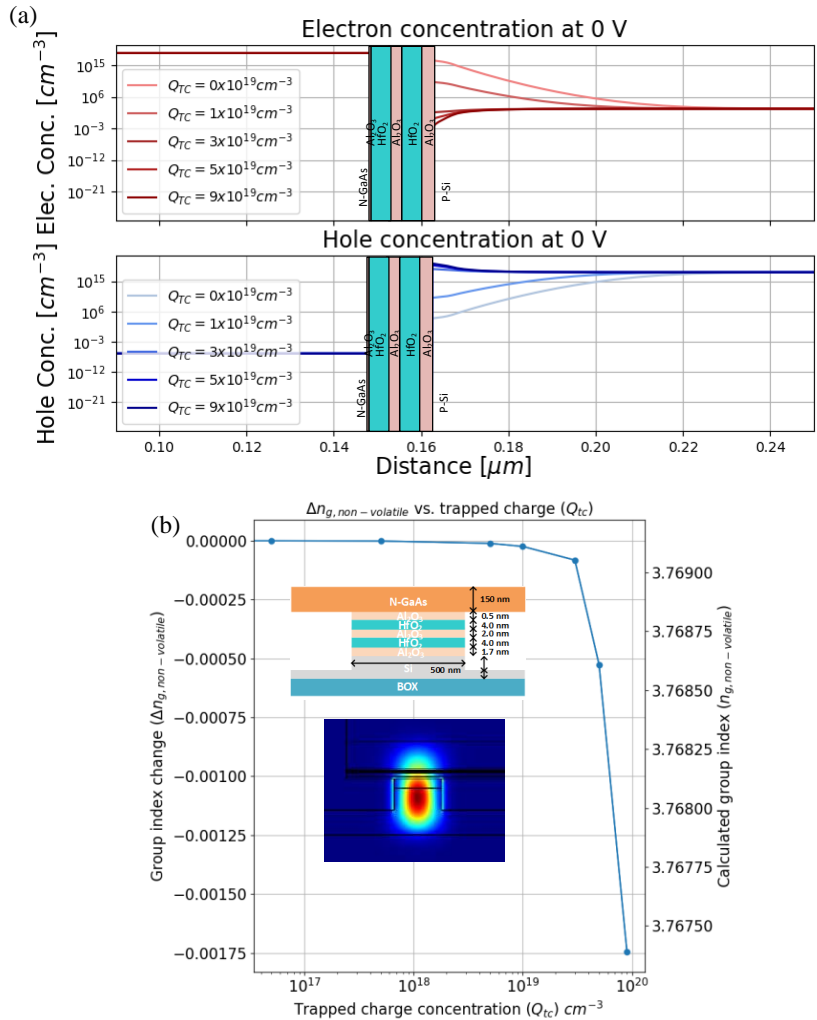

Fig. 10. Electron and hole concentrations for various QTC after set process (0 → -15 → 0 V) and (b) non-volatile group index change $\Delta n_{g,non\text{-}volatile}$ vs. $Q_{TC}$.

## Post-fabrication phase error correction

An alternative use of the non-volatile III-V/Si memristor cell is the role it can play in post-fabrication trimming or permanent phase error correction without consuming static electrical power. Inherent to the high index contrast of the silicon photonic material system, phase sensitive devices such as arrayed waveguide gratings (AWGs), lattice filters, and (de-)interleavers are sensitive to phase errors and are dependent on waveguide width, thickness, and refractive index non-homogeneity. In this case, the ability to use non-volatile phase tuning to correct the aforementioned errors is essential to minimizing power consumption. Changes in resonant the resonant wavelength can be described by the following equation [4]:

$$\Delta \lambda_0 = \left( \lambda_0 / n_g \right) \sqrt{\left( dn_{eff} / dw \cdot \Delta w \right)^2 + \left( dn_{eff} / dt \cdot \Delta t \right)^2} \qquad (12)$$

$\lambda_0, n_{eff}, n_g, \Delta w,$ and $\Delta t$ are the free-space wavelength, effective index, group index, width variation, and thickness variation respectively. Along with the group index $n_g = n_{eff} - \lambda_0 \cdot dn_{eff} / d\lambda$, the resonant wavelength variation for each dimension can be calculated as: $\Delta \lambda_0 / \Delta w = (\lambda_0 / n_g)(dn_{eff} / dw)$ and $\Delta \lambda_0 / \Delta t = (\lambda_0 / n_g)(dn_{eff} / dt)$. The effective index and group index as a function of width and thickness are plotted in Fig. *11*a – b. It can be seen that both values increase as waveguide dimensions increase because of increased modal confinement. Fig. *11*c – d illustrates width sensitivity ($dn_{eff}/dw$) and thickness sensitivity ($dn_{eff}/dt$). Throughout the paper, single-mode III-V/Si memristor waveguides are used and have design dimensions of height = 300 nm, width = 500nm, etch depth = 170nm, and GaAs thickness of 150 nm, thus resulting in effective index variations of $dn_{eff}/dw = 5.80 \times 10^{-4}$ /nm and $dn_{eff}/dt = -4.44 \times 10^{-5}$ /nm. Typically, the most critical parameter in controlling phase errors is the starting SOI wafer thickness uniformity.

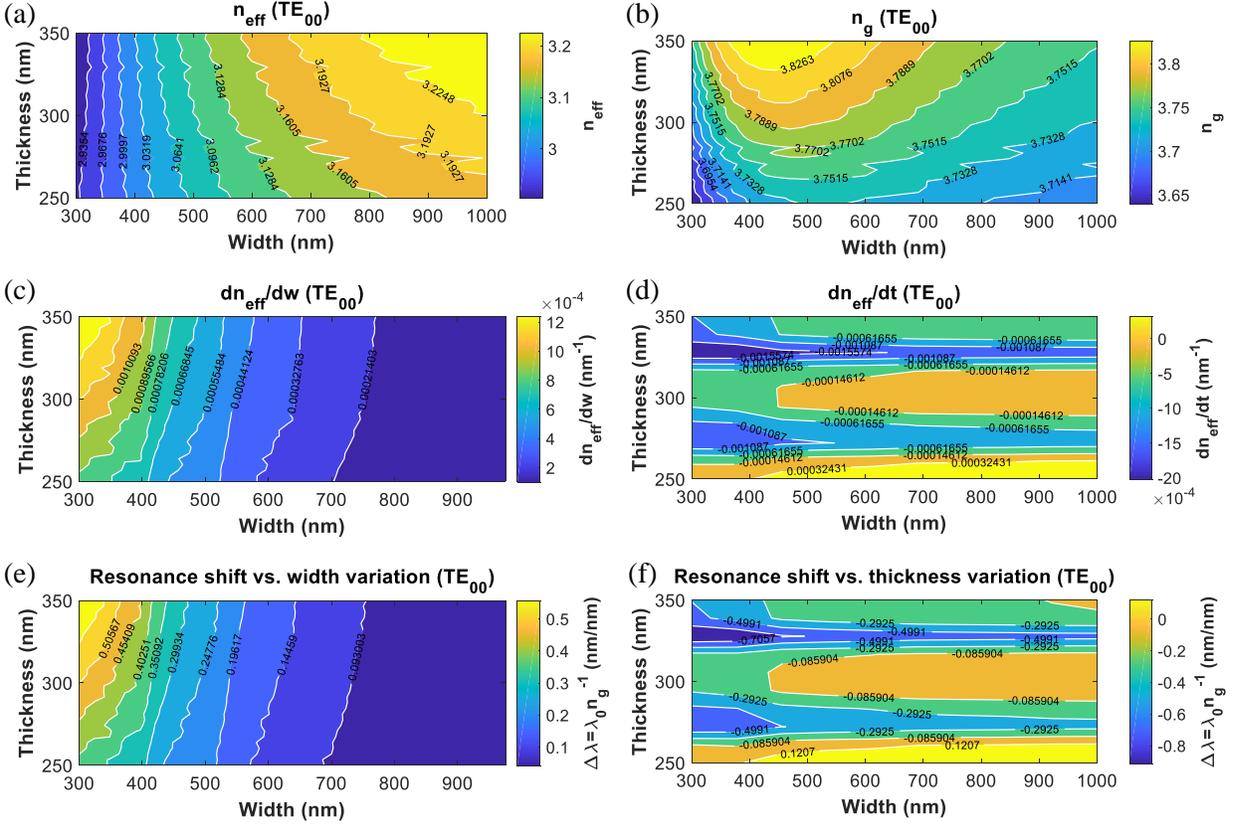

Fig. 11 III-V/Si memristor optical mode calculations for (a) effective index ($n_{eff}$), (b) group index ($n_g$), (c) effective index change vs. waveguide width ($dn_{eff}/dw$), (d) effective index change vs. waveguide thickness ($dn_{eff}/dt$), (e) wavelength shift ($\Delta\lambda_0/dw$) vs. waveguide width, (f) wavelength shift vs. waveguide thickness ($\Delta\lambda_0/dt$).

However, for the III-V/Si case, the GaAs thickness reduces this SOI thickness sensitivity significantly, and as a result, width variations play a larger role in phase errors by an order of magnitude. The wavelength shift variation are $\Delta\lambda_0/dw$ = 0.2457 nm/nm and $\Delta\lambda_0/dt$ = -0.0187 nm/nm as shown in Fig. *11*e – f. The use of wider waveguides can significantly reduce $\Delta\lambda_0/dw$ by another order of magnitude, however, the TE01 mode starts to appear at a width of 600 nm as seen in **Error! Reference source not found.**b. We believe, the III-V/Si CTM waveguide dimensions used throughout this paper offers the best design trade-off in terms of low $dn_{eff}/dw$ and $dn_{eff}/dt$ while maintaining single-mode operation. If we assume a charge trap concentration of $Q_{TC} > 9 \times 10^{19}$ cm$^{-3}$, which corresponds to a $\Delta n_{eff}^{III-V/Si} = 1.60 \times 10^{-3}$, the III-V/Si optical memristor may be capable of correcting a couple nm of variation in waveguide width and height.

## Supplementary Note 2: Fabrication

In-house device fabrications starts with a 100 mm SOI wafer which consists of a 350 nm thick top silicon layer and a 2 µm buried oxide (BOX) layer. The top silicon is thinned down to 300 nm by thermal oxidation and buffered hydrofluoric (HF) acid etching, thus leaving a clean silicon surface. Silicon waveguides are defined by a deep-UV (248 nm) lithography stepper and boron is implanted to create p ++ silicon contacts. Grating couplers, silicon rib waveguides, and vertical out-gassing channels (VOCs) are respectively patterned using the same deep-UV stepper and then subsequently etched 170 nm with Cl$_2$-based gas chemistry. Next, the silicon wafer goes through a Piranha clean followed by buffered hydrofluoric (HF) acid etching to remove any hard masks. Next, an oxygen plasma clean is performed followed by a SC1 and SC2 clean. The III-V wafer goes through a solvent clean consisting of acetone, methanol, and IPA, followed by oxygen plasma cleaning and a $NH_4OH:H_2O$ (1:10) dip for 1 minute. Next a dielectric of $Al_2O_3$ is deposited onto both GaAs and Si wafers via atomic layer deposition (ALD) by using 5 cycles of trimethylaluminum (TMA) + $H_2O$ with a target thickness of 0.5 nm on each side. All ALD deposition temperatures are performed at 300 °C. Next, a thickness target of 3 nm $HfO_2$ is deposited on each sample via 30 cycles of tetrakis (ethylmethylamino) hafnium (TEMAH) + $H_2O$. Finally, a thickness target of 1 nm $Al_2O_3$ is deposited on each sample via 10 cycles of TMA + $H_2O$. The two samples are then mated manually at room temperature and then wafer-bonded under pressure for 250 °C (2 hour ramp) for a total of 15 hours. After wafer-bonding, the backside of the III-V is mechanically lapped until ~ 100 μm of III-V is left. An $Al_{0.20}Ga_{0.80}As$ etch stop layer allows selective removal of the remaining GaAs substrate via wet etching ($H_2O_2:NH_4OH$ (30:1)). The $Al_{0.20}Ga_{0.80}As$ is finally removed in buffered hydrofluoric acid (BHF), thus leaving a 150 nm-thick n-GaAs thin film on top of the SOI substrate. A combination of Ge/Au/Ni/Au/Pd/Ti (400/400/240/4000/200/200 Å) is deposited onto the n-GaAs as an n-contact layer. Next, the III-V film is defined and dry etched for device regions that require the use of phase tuning. Metal contact with the p-Si consists of Ni/Ge/Au/Ni/Au/Ti (50/300/300/200/5000/200 Å). Next, a plasma enhanced chemical vapor deposition (PECVD) $SiO_2$ cladding is deposited and the vias are defined and etched. Finally, Ti/Au metal probe pads are defined to make contact with n-GaAs and p-Si layers. The relevant fabricated devices can be seen in Fig. *13*a-b.

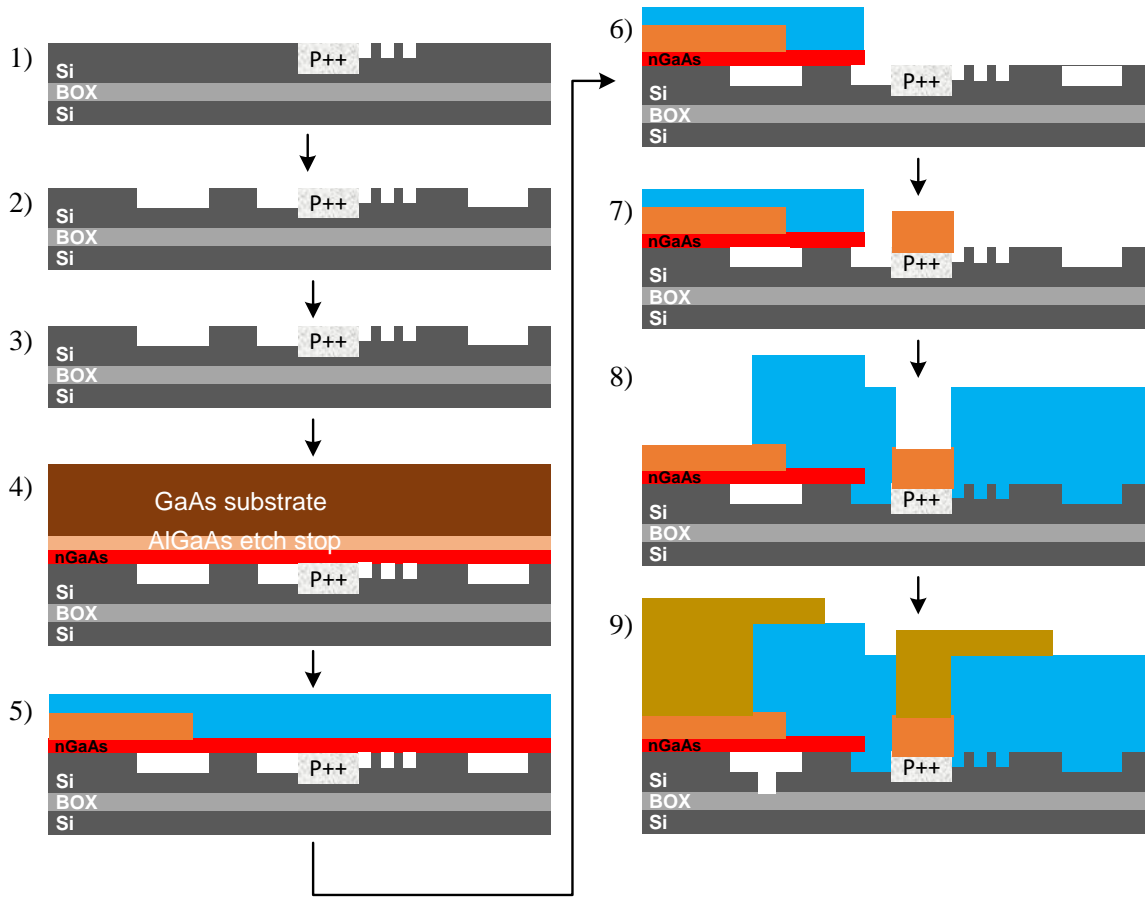

Fig. 12 Fabrication flow of heterogeneous III-V/Si photonic memristor devices.

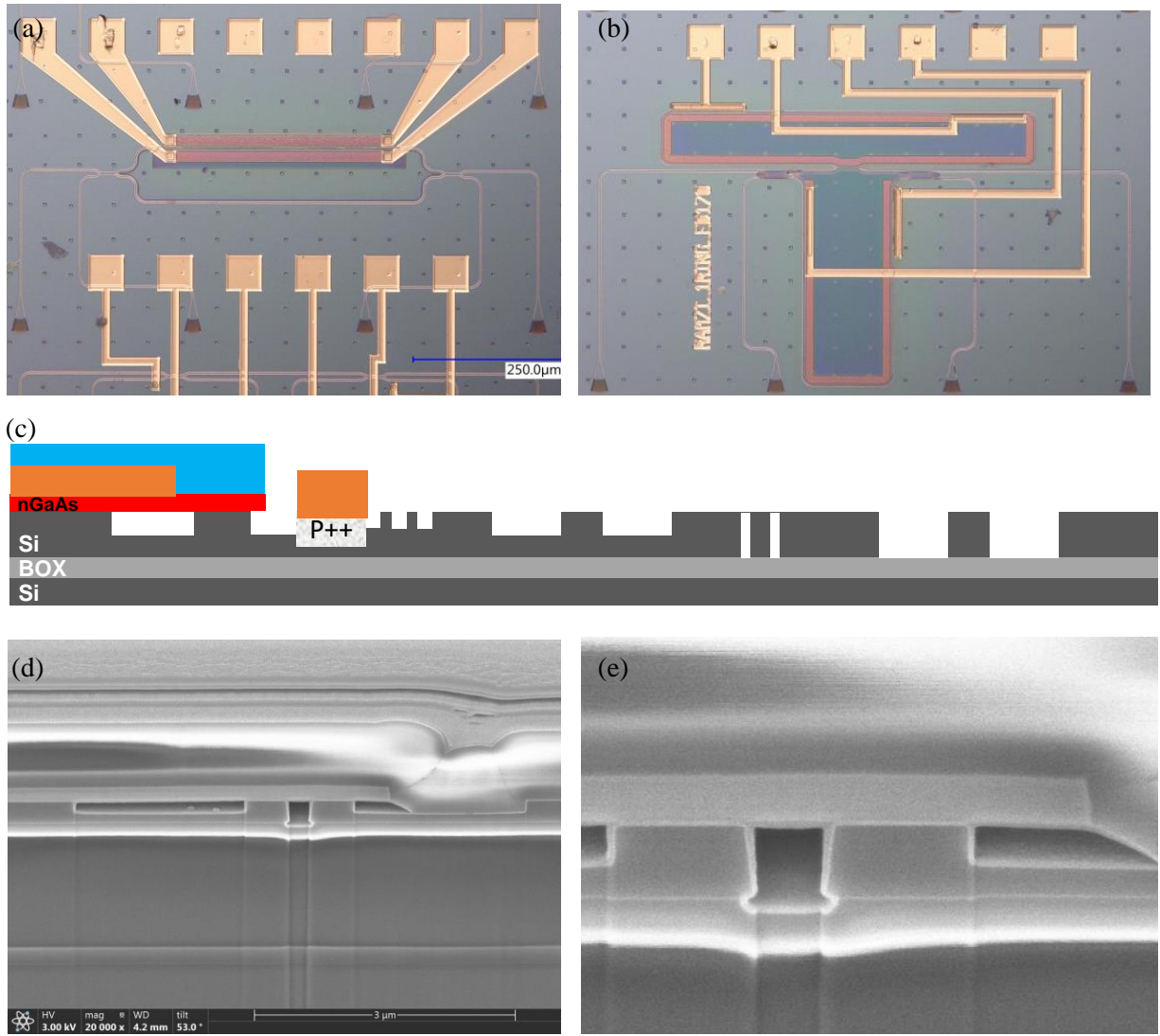

Fig. 13. Microscope images of fabricated III-V/Si photonic memristor devices (a) MZI, and (b) 1-RAMZI filter. (c) Schematic of 2-D cross-section. (d) - (e) SEM image of cross-section.

## Supplementary Note 3: Transmission electron microscope (TEM) measurements of initial, set, and reset states.

We performed high resolution transmission electron microscope (HRTEM) imaging of initial, set, and reset states as shown in *Fig. 14*a-c respectively. Each image is a separate sample that was biased and subsequently imaged. There does not appear to be any visual evidence of dielectric breakdown in the n-GaAs/$Al_2O_3$/$HfO_2$/$Al_2O_3$/$HfO_2$/ $Al_2O_3$/Si memristor structure. There are contrast differences and may indicate variability in sample preparation as well as imaging settings. The electron dispersive spectroscopy (EDS) line scans below the TEM images also indicate minimal atomic/interfacial changes. Fig. *15*a-c shows the 2D atomic composition mapping for each respective state. Fig. *16*a-c shows the quantitative 2D strain mapping of the oxide/Si interface for each respective state. It is observed that the set state has more stain locations compared to the initial and reset states.

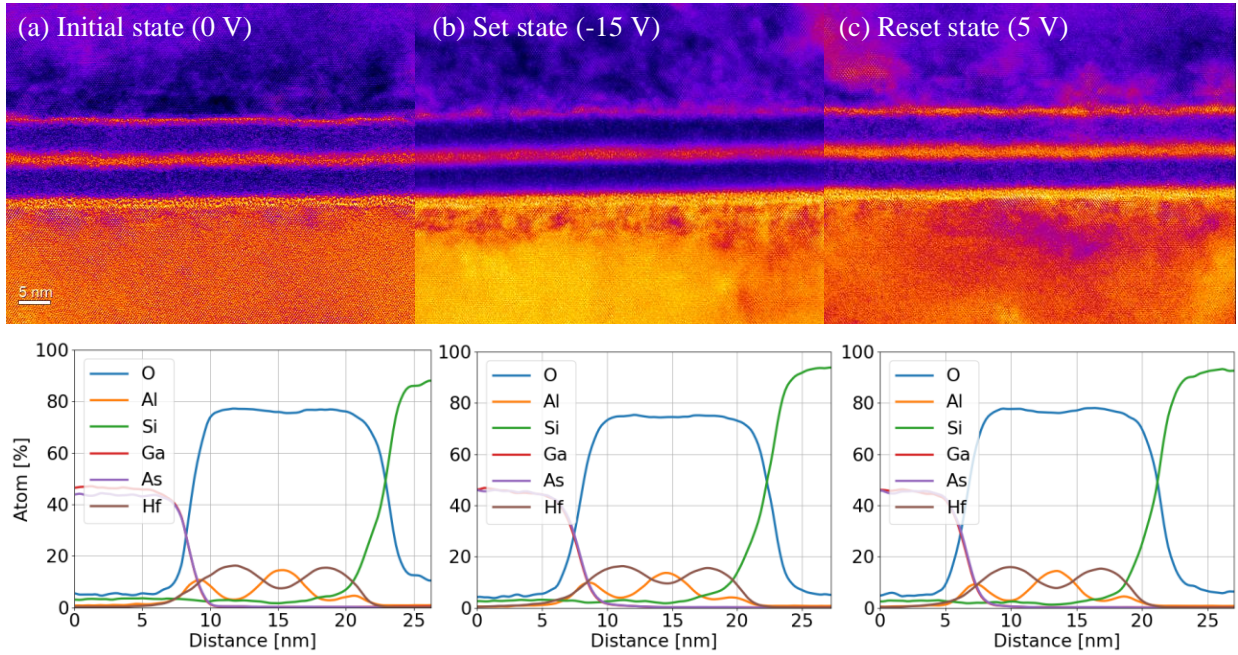

Fig. 14 TEM imaging and EDS line scan of n-GaAs/$Al_2O_3$/$HfO_2$/$Al_2O_3$/$HfO_2$/ $Al_2O_3$/Si memristor structure in (a) initial state (0 V), (b) set state (-15 V), and (c) reset state (+ 5 V).

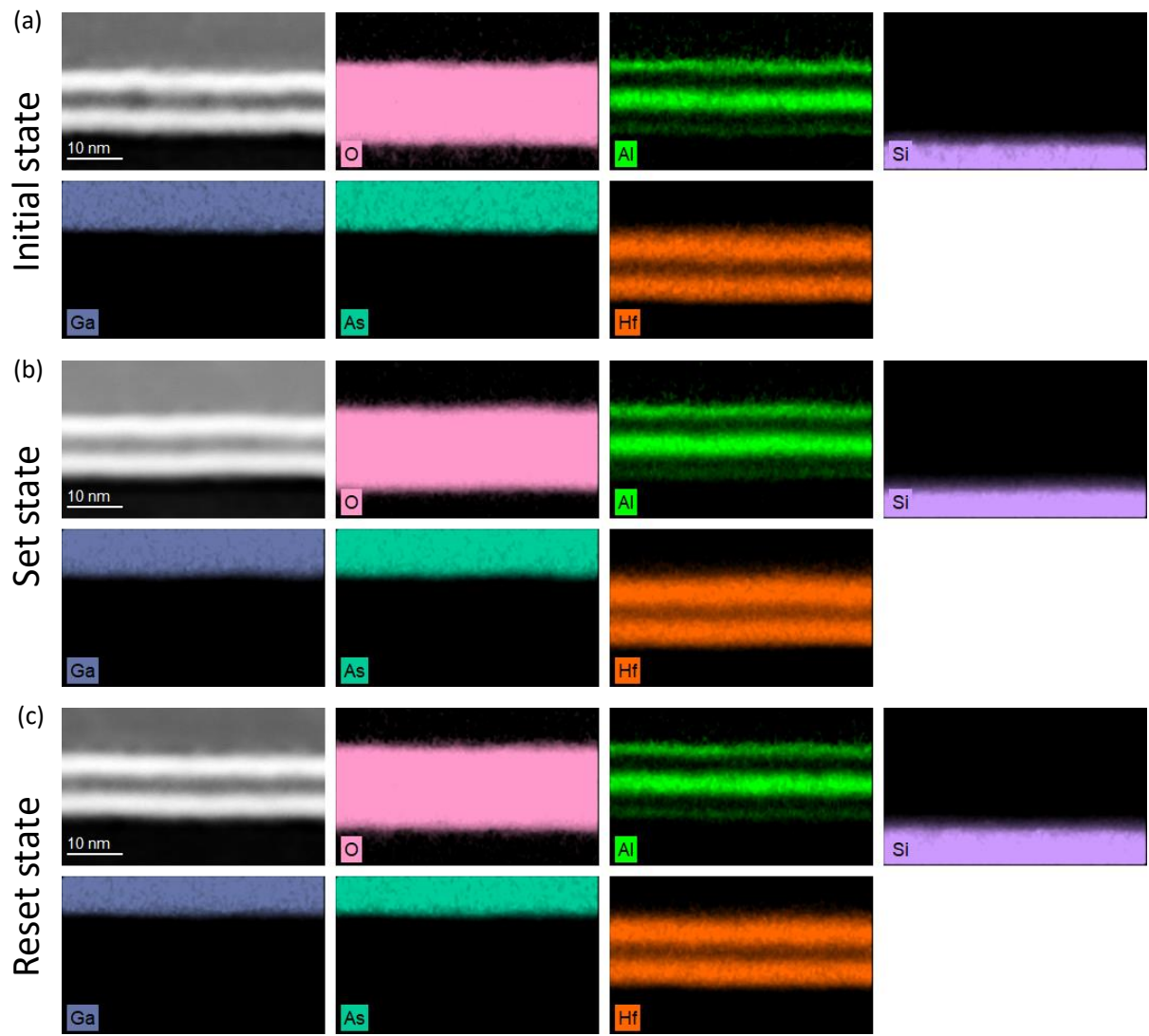

Fig. 15. 2D atomic composition mapping for (a) initial state (0 V), (b) set state (-15 V), and (c) reset state (+ 5 V).

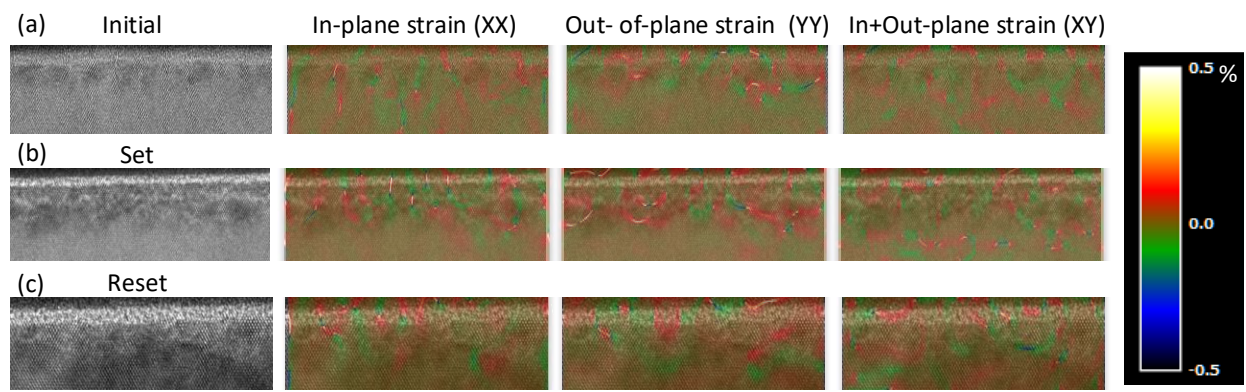

Fig. 16. 2D strain mapping via geometrical phase analysis (GPA) for (a) initial state (0 V), (b) set state (-15 V), and (c) reset state (+ 5 V).

# Supplementary Note 4: High speed $S_{21}$ measurements for multiple non-volatile states.

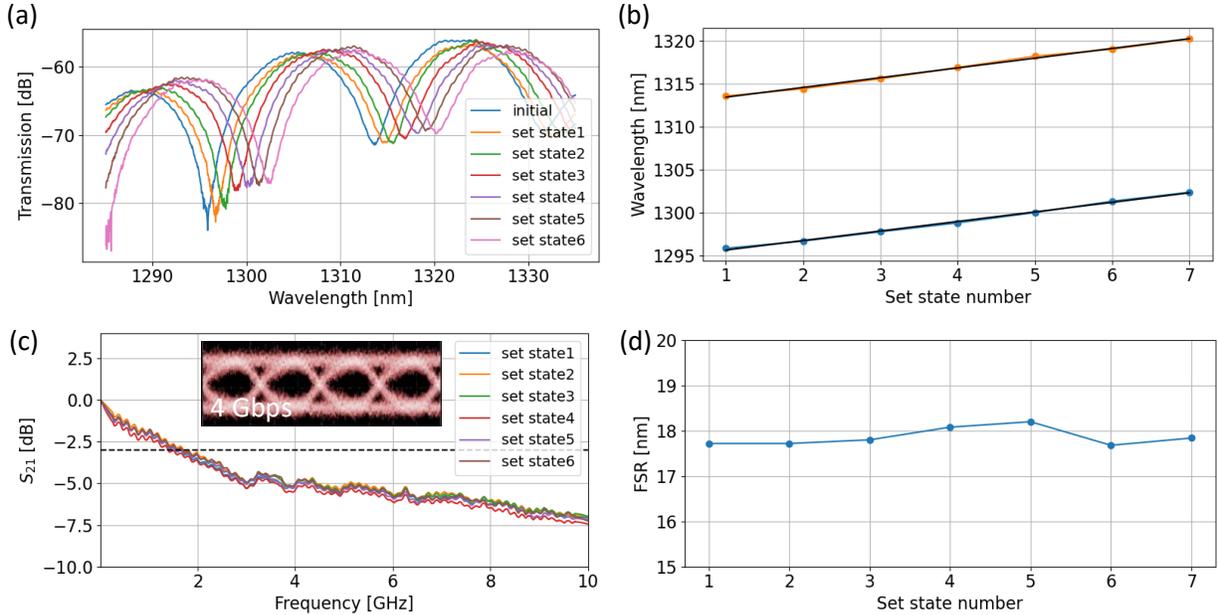

Fig. 17. (a) Optical spectrum of 6 non-volatile set-states, (b) corresponding wavelength shifts, (c) small-signal $S_{21}$ measurement for each set state ($f_{3dB}$ ~ 1.9 GHz) and an eye-diagram at 4 Gbps, and (d) extracted FSR for each state.

# Supplementary Note 5: An extreme case of phase shift tuning $> 4\pi$

Here is one particular device that exhibited a phase shift $\Delta\varphi > 4\pi$. This corresponds to a III-V/Si group index change of $\Delta n_g^{III-V/Si} = 13.7 \times 10^{-3}$ determined by the method mentioned in the main text. The same voltage cycling is performed (0 → -21 → 0 → 15 → 0 V) and an electrical and resistive hysteresis is observed in Fig. *18*a – b respectively. Applying a bias from 0 to -21 V results in a non-volatile wavelength shift of $\Delta\lambda_{non-volatile}$ ~ 40 nm wavelength shift with near negligible optical losses (Fig. *18*c). Ramping back down from -21 to 0 V does not shift the optical response back to the original state (Fig. *18*d) and has a non-volatile wavelength stability of ~ +/- 0.2 nm (35 GHz). This indicates a non-volatile phase shift of $\Delta\varphi > 4\pi$, at essentially 0 power consumption (recorded current = 10.4 pA at 0 V).

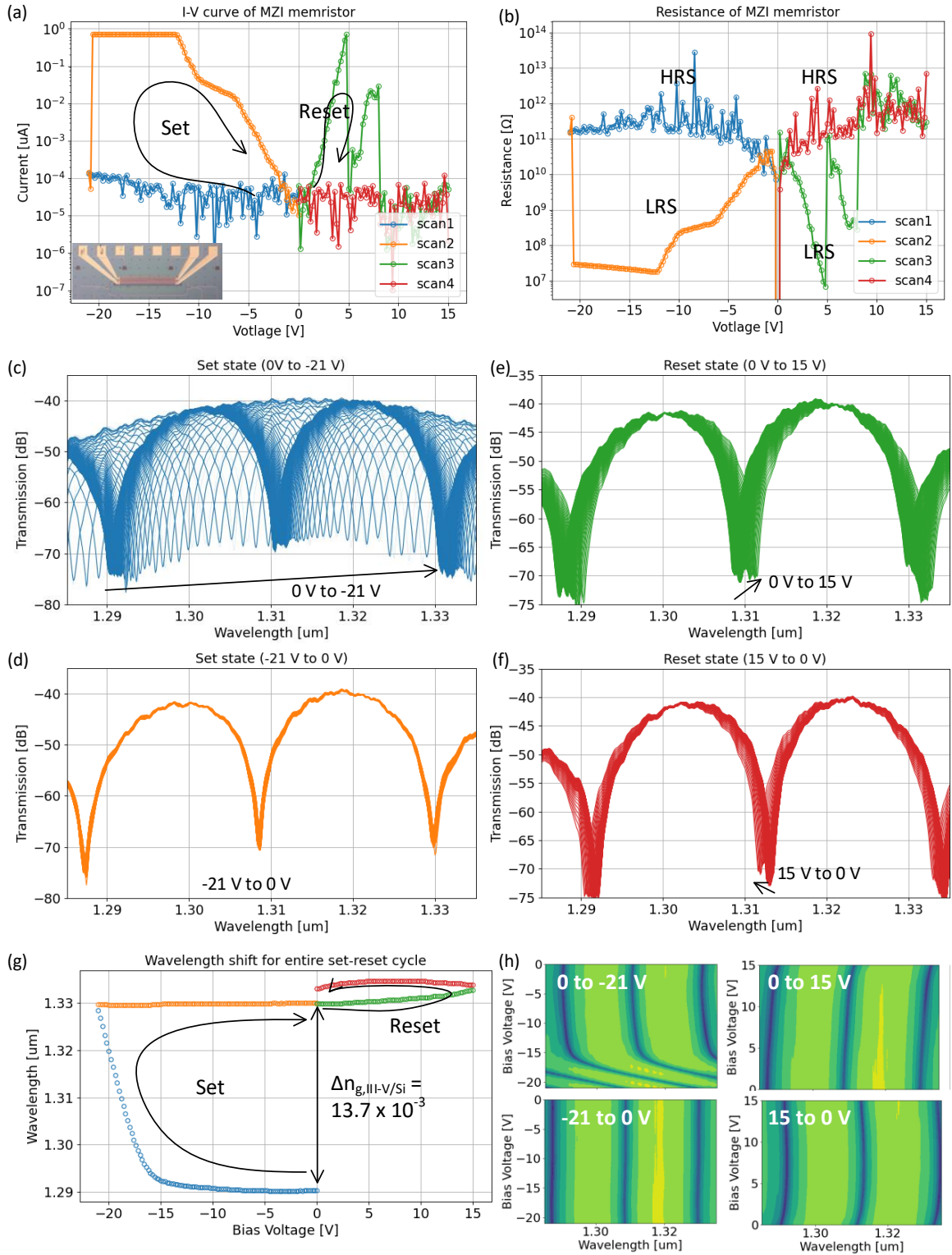

Fig. 18. (a) Measured I-V hysteresis indicating non-volatile memristive set/reset states. (b) Corresponding resistance indicating regions of high-resistance-states (HRS) and low-resistance-states (LRS). Measured optical spectrum for (c) "set" (0 to -21 V), (d) turning off "set" (-21 to 0V), (e) "reset" (0 to 15 V), and (f) turning off "reset" (15 to 0 V). (g) Tracked resonance vs. voltage, (h) spectral evolution vs. voltage.